\documentclass[11pt]{article}
\usepackage{amsmath}
\usepackage{amssymb}
\usepackage{times}
\usepackage{color}
\usepackage{epsfig,graphics}
\usepackage{bm}
\usepackage{mathrsfs}
\usepackage{enumerate}
\usepackage{authblk}
\usepackage{indentfirst}
\begin{document}

\pagestyle{plain}
\hsize = 6.5 in
\vsize = 8.5 in
\hoffset = -0.75 in
\voffset = -0.5 in
\baselineskip = 0.29 in

\newcommand{\DF}[2]{{\displaystyle\frac{#1}{#2}}}

\title{Information Landscape and Flux, Mutual Information Rate Decomposition and Entropy Production}

\author[1]{Qian Zeng}
\author[1,2]{Jin Wang \thanks{To whom correspondence should be addressed. E-mail: jin.wang.1@stonybrook.edu}}
\affil[1]{State Key Laboratory of Electroanalytical Chemistry, Changchun Institute of Applied Chemistry, Changchun, Jilin 130022, China}
\affil[2]{Department of Chemistry and Physics, State University of New York, Stony Brook, NY 11794 USA.}


\date{\today}

\maketitle

\begin{abstract}
We explore the dynamics of information systems. We show that the driving force for information dynamics is determined by both the information landscape and information flux which determines the equilibrium time reversible and the nonequilibrium time-irreversible behaviours of the system respectively. We further demonstrate that the mutual information rate between the two subsystems can be decomposed into the time-reversible and time-irreversible parts respectively, analogous to the information landscape-flux decomposition for dynamics. Finally, we uncover the intimate relation between the nonequilibrium thermodynamics in terms of the entropy production rates and the time-irreversible part of the mutual information rate. We demonstrate the above features by the dynamics of a bivariate Markov chain.
\end{abstract}

\section{Introduction}

There are growing interests in studying the information systems in the fields of control theory, information theory, communication theory, and biophysics \cite{0, 1,2,3,4,5}. Significant progresses have been made recently towards the understanding of the information system in terms of information thermodynamics \cite{9, 10, 11, 12}. However, the identification of the global driving force for the information system dynamics is still challenging. Here we would like to fill the gap by quantifying the driving forces for the information system dynamics. Inspired by the recent development of landscape and flux theory for the non-equilibrium systems \cite{13, 14, 15}, we will show that the driving force for information dynamics is determined by both the information landscape and information flux. The information flux is a measure of the degree of nonequilibirumness or time irreversibility. Mutual information represents the correlation between two information subsystems. We uncovered that the mutual information rate between the two subsystems can be decomposed into the time-reversible and time-irreversible parts respectively. This is originated from the information landscape-flux decomposition for dynamics. An important signature of nonequilibriumness is the entropy production or energy cost. We also uncover the intimate relation between the entropy production rates and the time-irreversible part of the mutual information rate. We demonstrate the above features by the dynamics of a bivariate Markov chain.

\section{Bivariate Markov Chains}

Markov chains have been often assumed for the underlying information dynamics of the total system in random environments. That is, the two subsystems together forms a Markov chain in continuous or discrete times, which is the so-called \emph{Bivariate Markov Chain}(BMC). The processes of the two subsystems are correspondingly said to be marginal processes or marginal chain. The BMC was used to model ion channel currents \cite{1}, it was also used to model delays and congestion in a computer network \cite{2}. Recently, different models of BMC appeared in non-equilibrium statistical physics for capturing or implementing the Maxwell's demon \cite{3,4,5}, which can be seen as one marginal chain in the BMC playing feedback control to the other marginal chain. Although the BMC has been studied for decades, there are still challenges on quantifying the  dynamics of the whole as well as the two subsystems. This is because neither of them needs to be Markovian chain in general \cite{6}, and the quantifications of the probabilities (densities) for the trajectories of the two subsystems involve complex random matrices manipulations \cite{7}. This leads to the problem not exactly analytically solvable. The corresponding numerical solutions often lack direct mathematical and physical interpretations.

The conventional analysis of the BMC focuses on the mutual information \cite{8} of the two subsystems for quantifying the underlying information correlations. There are three main representations on this. The first one was proposed by Sagawa \cite{9,10} for explaining the mechanism of Maxwell's demon in Szilard's engine. In this representation, the mutual information between the demon and controlled system characterizes the observation and the feedback of the demon. This leads to an elegant way which includes the increment of the mutual information into a unified fluctuation relation. The second representation was proposed by Esposito \cite{11} in an attempt to explain the violation of the second law in a specified BMC, the bipartite model, where the mutual information is divided into two parts corresponding to the two subsystems respectively, which were said to be the information flows. This representation tries to explain the mechanism of the demon because one can see that the information flows do contribute to the entropy production to both demon and controlled system. The first two representations are based on the ensembles of the subsystem states. This means that the mutual information is defined only on the time-sliced distributions of the system states, which somehow lacks the information of subsystem dynamics: the time-correlations of the observation and feedback of the demon. The last representation was seen in the work of Seifert \cite{12} where he used a more general definition of mutual information in information theory, which is defined on the trajectories of the two subsystem. More exactly, this is the so-called \emph{Mutual Information Rate} (MIR) which quantifies the correlation between the two subsystem dynamics. However, due to the difficulties from the possible underlying non-Markovian property of the marginal chains, exactly solvable models and comprehensive conclusions are still challenging from this representation.

In this study, we study the discrete-time BMC in both stochastic dynamics. To avoid the technical difficulty caused by non-Markovian dynamics, we first assume that the two marginal chains follow the Markovian dynamics. We explore the time-irreversibility of BMC and marginal processes in steady state. Then we decompose driving force for the underlying information dynamics  as the  information landscape and information flux  \cite{13,14,15} representing the time-reversible parts and time-irreversible parts respectively. We also prove that the non-vanishing flux fully describes the time-irreversibility of BMC and marginal processes.

We focus on the mutual information rate between the two marginal chains in information dynamics. Since the two marginal chains are assumed to be Markov chains here, the mutual information rate is exactly analytically solvable, which can be seen as the averaged conditional correlation between the two subsystem states. Here the conditional correlations reveal the time correlations between the past states and the future states.

Corresponding to the landscape-flux decomposition in stochastic dynamics, we decompose the MIR into two parts: the time-reversible and time-irreversible parts respectively. The time-reversible part measures the part of the correlations between the two marginal chains in both forward and backward processes of BMC. The time-irreversible part measures the difference between the correlations in forward and backward processes of BMC respectively. We can see that a non-vanishing time-irreversible part of the MIR must be driven by a non-vanishing flux in steady state, and can be seen as the sufficient condition for a BMC to be time-irreversible.

We also reveal the important fact that the time-irreversible parts of MIR contributes to the nonequilibrium \emph{Entropy Production Rate} (EPR) of the BMC by the simple equality:
\begin{eqnarray*}
  \text{EPR of BMC} = \text{EPR of 1st marginal chain} + \text{EPR of 2nd marginal chain} + 2\times\text{time-irreversible part of MIR}.
\end{eqnarray*}
And this relation may help to develop general theory on nonequilibrium interacting information system dynamics.

\section{ Information Landscape and Information Flux for Determining the Information Dynamics, Time-Irreversibility}

Consider a finite-state, discrete-time, ergodic, and irreducible bivariate Markov chain
\begin{eqnarray}
  Z=(X,S)=\{(X(t),S(t)),t\ge 1\}.
\end{eqnarray}
We assume that the state space of $X$ is given by $\mathcal{X}=\{1,...,d\}$ and the state space of $S$ is given by $\mathcal{S}=\{1,...,l\}$. The state space of $Z$ is then given by $\mathcal{Z}=\mathcal{X}\times\mathcal{S}$. The time evolution of distribution of $Z$ is characterized by the following master equation in discrete time,
\begin{eqnarray}
  p_z(z;t+1)=\sum_{z'}q_z(z|z')p_z(z';t),\ \text{for } t\ge 1, \text{ and }z\in \mathcal{Z}
\end{eqnarray}
where $p_z(z;t)=p_z(x,s;t)$ is the probability of observing state $z$ (or joint probability of $X=x$ and $S=s$) at time $t$; $q_z(z|z')=q_z(x,s|x',s')\ge 0$ are the transition probabilities from $z'=(x',s')$ to $z=(x,s)$ respectively and are with $\sum_z q_z(z|z')=1$.

We assume that there exists a unique stationary distribution $\pi_z$ such that $\pi_z(z)=\sum_{z'}q_z(z|z')\pi_z(z')$. Then given arbitrary initial distribution, the distribution goes to $\pi_z$ exponentially fast in time. If the initial distribution is $\pi_z$, we say that $Z$ is in \emph{Steady State} (SS) and our discussion is based on this SS.

The marginal chains of $Z$, i.e., $X$ and $S$, do not need to be Markov chains in general. For simplicity of analysis, we assume that both marginal chains are Markov chains and the corresponding transition probabilities are given by $q_x(x|x')$ and $q_s(s|s')$ (for $x,x'\in\mathcal{X}$ and $s,s'\in\mathcal{S}$) respectively. Then we have the following master equations for $X$ and $S$,
\begin{eqnarray}
  p_x(x;t+1)=\sum_{x'}q_x(x|x')p_x(x';t),
\end{eqnarray}
and
\begin{eqnarray}
  p_s(s;t+1)=\sum_{s'}q_s(s|s')p_s(s';t),
\end{eqnarray}
where $p_x(x;t)$ and $p_s(s;t)$ are the probabilities of observing $X=x$ and $S=s$ at time $t$ respectively.

We consider that both Eqs.(3,4) have unique stationary solutions $\pi_x$ and $\pi_s$ which satisfy $\pi_x(x)=\sum_{x'} q_x(x|x')\pi_x(x')$ and $\pi_s(s)=\sum_{s'}q_s(s|s')\pi_s(s')$ respectively. Also, we assume that when $Z$ is in SS, $\pi_x$ and $\pi_s$ are also achieved. The relations between $\pi_x$, $\pi_s$ and $\pi_z$ read,
\begin{eqnarray}
  \begin{cases}
    \pi_x(x)=\sum_s \pi_z(x,s),\\
    \pi_s(s)=\sum_x \pi_z(x,s).
  \end{cases}
\end{eqnarray}

In the rest of this paper, we let $X^T=\{X(1),X(2),...,X(T)\}$, $S^T=\{S(1),S(2),...,S(T)\}$, and $Z^T=\{Z(1),Z(2),...,Z(T)\}=(X^T,S^T)$ denote the time sequences of $X$, $S$, and $Z$ in time $T$ respectively.

To characterize the time-irreversibility of the Markov chain $C$ in stochastic dynamics in SS, we introduce the concept of probability flux. Here we let $C$ denote arbitrary Markov chain in $\{Z,X,S\}$, and let $c$, $\pi_c$, $q_c$, and $C^T$ denote arbitrary state of $C$, the stationary distribution of $C$, the transition probabilities of $C$, and a time sequence of $C$ in time $T$ and in SS, respectively.

The averaged number transitions from the state $c'$ to state $c$, denoted by $N(c'\to c)$, in unit time in SS can be obtained as
\begin{eqnarray*}
  N(c'\to c)=\pi_c(c')q_c(c|c').
\end{eqnarray*}
This is also the probability of the time sequence $C^T=\{C(1)=c',C(2)=c\}$, ($T=2$). Correspondingly, the averaged number of reverse transitions, denoted by $N(c\to c')$, reads
\begin{eqnarray*}
  N(c\to c')=\pi_c(c)q_c(c'|c).
\end{eqnarray*}
This is also the the probability of the time-reverse sequence $\widetilde{C}^T=\{C(1)=c,C(2)=c'\}$, ($T=2$). The difference between these two transition numbers measures the time-reversibility of the forward sequence $C^T$ in SS,
\begin{eqnarray}
  J_c(c'\to c)&=&N(c'\to c)-N(c\to c')\nonumber\\
  &=&P(C^T)-P(\widetilde{C}^T)\nonumber\\
  &=&\pi_c(c')q_c(c|c')-\pi_c(c)q_c(c'|c), \text{ for }C=X,S,\text{ or }Z.
\end{eqnarray}
Then, $J_c(c'\to c)$ is said to be the probability flux from $c'$ to $c$ in SS. If $J_c(c'\to c)=0$ for arbitrary $c'$ and $c$, then $C^T$ ($T=2$) is time-reversible; otherwise when $J_c(c'\to c)\neq 0$, $C^T$ is time-irreversible. Clearly, we have from Eq. (6) that
\begin{eqnarray}
  J_c(c'\to c)=-J_c(c\to c').
\end{eqnarray}

The transition probability determines the evolution dynamics of the information system. We can decompose the transition probabilities $q_c(c|c')$ into two parts: the time-reversible part $D_c$ and time-irreversible part $B_c$, which read
\begin{eqnarray}
&&q_c(c|c')=D_c(c'\to c)+B_c(c'\to c),\ \text{with}\\
  &&\begin{cases}\nonumber
    D_c(c'\to c)=\frac{1}{2\pi_c(c')}(\pi_c(c')q_c(c|c')+\pi_c(c)q_c(c'|c)),\\
    B_c(c'\to c)=\frac{1}{2\pi_c(c')}J_c(c'\to c).
  \end{cases}
\end{eqnarray}

From this decomposition, we can see that the information dynamics is determined by two driving forces. One of the driving force is determined by the steady state probability distribution and is time reversible. The other driving force for the information system dynamics is the steady state probability flux which breaks the detailed balance and quantify the time irreversibility. Since the steady state probability measures the weight of the information state, therefore it quantifies the information landscape. If we define the potential landscape for the information system as $\phi=-\log \pi$, then the $D_c(c'\to c)=\frac{1}{2}(q_c(c|c')+\frac{\pi_c(c)}{\pi_c(c')} q_c(c'|c))=\frac{1}{2}(q_c(c|c')+ \exp[-(\phi_c(c)-\phi_c(c')] q_c(c'|c))$ becomes the difference or "gradient" in the potential landscape. Therefore, this reversible part of the information dynamics is determined by the "gradient" of the information landscape. The steady state probability flux measures the information flow in the dynamics and therefore can be termed as the information flux. It is a direct measure of the nonequilibriumness in terms of time irreversibility.

By Eqs.(7,8), we have the following relations
\begin{eqnarray}
  \begin{cases}
  \pi_c(c')D_c(c'\to c)=\pi_c(c)D_c(c\to c'),\\
  \pi_c(c')B_c(c'\to c)=-\pi_c(c)B_c(c\to c').
  \end{cases}
\end{eqnarray}
As we can see in next section, $D_c$ and $B_c$ are useful for us to quantify time-reversible and time-irreversible observables of $C$ respectively.

We give the interpretation that the non-vanishing probability flux $J_c$ fully measures the time-irreversibility of the chain $C$ in time $T$ for $T\ge 2$.
Let $C^T$ be arbitrary sequence of $C$ in SS, and with no loss of generality we let $T=3$. Similar to Eq. (6), the measure of time-irreversibility of $C^T$ can be given by the difference between the probability of $C^T=\{C(1),C(2),C(3)\}$ and that of its time-reversal $\widetilde{C}^T=\{C(3),C(2),C(1)\}$, such as
\begin{eqnarray*}
&&P(C^T)-P(\widetilde{C}^T)\nonumber\\
&&=\pi_c(C(1))q_c(C(2)|C(1))q_c(C(3)|C(2))-\pi_c(C(3))q_c(C(2)|C(3))q_c(C(1)|C(2))\nonumber\\
&&=\pi_c(C(1))\left(D_c(C(1)\to C(2))+B_c(C(1)\to C(2))\right)\left(D_c(C(2)\to C(3))+B_c(C(2)\to C(3))\right)-\nonumber\\
& \ \ &\pi_c(C(3))\left(D_c(C(3)\to C(2))+B_c(C(3)\to C(2))\right)\left(D_c(C(2)\to C(1))+B_c(C(2)\to C(1))\right),\nonumber\\
& \ \ \ \ \ \ \ \ \ \ \ \ \ \ \ &\text{          for } C=X,S \text{ or }Z.
\end{eqnarray*}
Then by the relations given in Eq.(9), we have $P(C^T)-P(\widetilde{C}^T)=0$ holds for arbitrary $C^T$ if and only if $B_c(C(1)\to C(2))=B_c(C(2)\to C(3))=0$ or equivalently $J_c(C(1)\to C(2))=J_c(C(2)\to C(3))=0$. This conclusion can be made for arbitrary $T>3$. Thus, non-vanishing $J_c$ can fully describe the time-irreversibility of $C$ for $C=X,S$, or $Z$.

We show the relations between the fluxes of the whole system $J_z$ and of the subsystem $J_x$ as following:
\begin{eqnarray}
  J_x(x'\to x)&=&\pi_x(x')q_x(x|x')-\pi_x(x)q_x(x'|x)\nonumber\\
  &=&P(\{x',x\})-P(\{x,x'\})\nonumber\\
  &=&\sum_{s,s'}\left(P(\{(x',s'),(x,s)\})-P(\{(x,s),(x',s')\})\right)\nonumber\\
  &=&\sum_{s,s'}\left(\pi_z(x',s')q_z(x,s|x',s')-\pi_z(x,s)q_z(x',s'|x,s)\right)\nonumber\\
  &=&\sum_{s,s'}J_z((x',s')\to(x,s)).
\end{eqnarray}
Similarly, we have
\begin{eqnarray}
J_s(s'\to s)=\sum_{x,x'}J_z((x',s')\to(x,s)).
\end{eqnarray}
These relations indicate that the subsystem fluxes $J_x$ and $J_s$ can be seen as the coarse-grained levels of total system flux $J_z$ by averaging over the other part of the system $S$ and $X$ respectively. We should emphasize that, Non-vanishing $J_z$ does not mean $X$ or $S$ is time-irreversible and vice versa.

\section{Mutual Information Decomposition to Time-Reversible and Time-Irreversible Parts}

According to the information theory, the two interacting information systems represented by bivariate Markov chain $Z$ can be characterized by the \emph{Mutual Information Rate} (MIR) between the marginal chains $X$ and $S$ in SS. The mutual information rates represents correlation between two interacting infomration systems. The MIR is defined on the probabilities of all possible time sequences, $P(Z^T)$, $P(X^T)$, and $P(S^T)$, and is given by
\begin{eqnarray}
  I(X,S)=\lim_{T\to \infty}\frac{1}{n}\sum_{Z^T}P(Z^T)\log\frac{P(Z^T)}{P(X^T)P(S^T)}.
\end{eqnarray}
It measures the correlation between $X$ and $S$ in unit time, or say, the efficient bits of information that $X$ and $S$ exchange with each other in unit time. The MIR must be non-negative, and a vanishing $I(X,S)$ indicates that $X$ and $S$ are independent of each other. More explicitly, the corresponding probabilities of these sequences can be evaluated by using Eqs.(2,3,4), we have
\begin{eqnarray*}
  \begin{cases}
    P(X^T)=\pi_x(X(1))\prod_{t=1}^{T-1}q_x(X(t+1)|X(t)),\\
    P(S^T)=\pi_s(S(1))\prod_{t=1}^{T-1}q_s(S(t+1)|S(t)),\\
    P(Z^T)=\pi_z(Z(1))\prod_{t=1}^{T-1}q_z(Z(t+1)|Z(t)).
  \end{cases}
\end{eqnarray*}
By substituting these probabilities into Eq.(12) (see Appendix), we have the exact expression of MIR as
\begin{eqnarray}
  I(X,S)&=&\sum_{z,z'}\pi_z(z')q_z(z|z')\log\frac{q_z(z|z')}{q_x(x|x')q_s(s|s')}\nonumber\\
  &=&\big\langle i(z|z')\rangle_{z',z}\ge 0, \text{ for }z=(x,s),\text{ and }z'=(x',s').
\end{eqnarray}
where $i(z|z')=\log\frac{q_z(z|z')}{q_x(x|x')q_s(s|s')}$ is the conditional (Markovian) correlation between the states $x$ and $s$ when the transition $z'=(x',s')\to z=(x,s)$ occurs. This indicates that when the two marginal processes are both Markov, the MIR is the average of the conditional (Markovian) correlations. These correlations are measurable when transitions occur and can be seen from the observables of $Z$.

By noting the decomposition of transition probabilities in Eq. (8), we have a corresponding decomposition of $I(X,S)$ such as
\begin{eqnarray}
&&I(X,S)=I_D(X,S)+I_B(X,S),\ \text{with}\\
  &&\begin{cases}\nonumber
    I_D(X,S)=\sum_{z,z'}\pi_z(z')D_z(z|z')i(z|z')=\frac{1}{2}\sum_{z,z'} (\pi_z(z')q_z(z|z')+\pi_z(z)q_z(z'|z))i(z|z'),\\
    I_B(X,S)=\sum_{z,z'}\pi_z(z')B_z(z|z')i(z|z')=\frac{1}{2}\sum_{z,z'}J_z(z|z')i(z|z')=\frac{1}{4}\sum_{z,z'}J_z(z|z')(i(z|z')-i(z'|z)).
  \end{cases}
\end{eqnarray}
This means that the mutual information representing the correlations between the two interacting systems can be decomposed into time reversible equilibrium part and time irreversible nonequilibrium part. The origin of this is from the fact the underlying information dynamics is determined by both the time reversible information landscape and time irreversible information flux. These equations are very important to establish the link to the time-irreversibility. We now give further interpretation for  $I_D(X,S)$ and $I_B(X,S)$:

Consider a bivariate Markov chain $Z$ in SS wherein $X$ and $S$ are dependent of each other, i.e., $I(X,S)=I_D(X,S)+I_B(X,S)>0$. By the ergodicity of $Z$, we have the MIR which measures the averaged conditional correlation along the time sequences $Z^T$,
\begin{eqnarray*}
 \lim_{T\to \infty}\frac{1}{T}\big\langle i(Z(t+1)|Z(t))\rangle_{Z^T} = I(X,S), \text{ for } 1<t<T.
\end{eqnarray*}
Then $I_B(X,S)$ measures the change of averaged conditional correlation between $X$ and $S$ when a sequence of $Z$ turns back in time,
\begin{eqnarray*}
  \lim_{T\to \infty}\frac{1}{T}\big\langle i(Z(t+1)|Z(t))-i(Z(t)|Z(t+1))\big\rangle_{Z^T} = 2 I_B(X,S).
\end{eqnarray*}
A negative $I_B(X,S)$ shows that the correlation between $X$ and $S$ becomes strong in the time-reversal process of $Z$; A positive $I_B(X,S)$ shows that the correlation becomes weak in the time-reversal process of $Z$. Both two cases show that the $Z$ is time-irreversible since we have a non-vanishing $J_z$. But the case of $I_B(X,S)=0$ is complicated, since it indicates either a vanishing $J_z$ or a non-vanishing $J_z$. Anyway, we see that a non-vanishing $I_B(X,S)$ is a sufficient condition for $Z$ to be time-irreversible. On the other hand, $I_D(X,S)=I(X,S)-I_B(X,S)$ measures the correlation remaining in the backward process of $Z$.

\section{Relationship Between Mutual Information and Entropy Production}

The \emph{Entropy Production Rates} (EPR) at steady state is a quantitative nonequilibriumness measure which characterizes the time-irreversibility of the underlying processes. The EPRs of the information system described by the bivariate Markov chains here can be given by
\begin{eqnarray}
  \begin{cases}
    R_z=\frac{1}{2}\sum_{z,z'}J_z(z'\to z)\log\frac{q_z(z|z')}{q_z(z'|z)}\ge0,\\
    R_x=\frac{1}{2}\sum_{x,x'}J_x(x'\to x)\log\frac{q_x(x|x')}{q_x(x'|x)}\ge0,\\
    R_s=\frac{1}{2}\sum_{s,s'}J_s(s'\to s)\log\frac{q_s(s|s')}{q_s(s'|s)}\ge0,\\
  \end{cases}
\end{eqnarray}
where total and subsystem entropy productions $R_z$, $R_x$, and $R_s$ correspond to $Z$, $X$, and $S$ respectively. Here, $R_z$ usually contains the detailed interaction information of the system (or subsystems) and environments; $R_x$ and $R_s$ provide the coarse-grained information of time-irreversible observables of $X$ and $Z$ respectively. Each non-vanishing EPR indicates that the corresponding Markov chain is time-irreversible. Again, we emphasize that a non-vanishing $R_z$ does not mean $X$ or $S$ is time-irreversible and vice versa.

We are interested in the connection between these EPRs and mutual information. We can associate them with $I_B(X,S)$ by noting Eqs.(10,11,14). We have
\begin{eqnarray}
 I_B(X,S)&=&\frac{1}{4}\sum_{z,z'}J_z(z|z')(i(z|z')-i(z'|z))\nonumber\\
 &=&\frac{1}{4}\sum_{z,z'}J_z(z|z')\log\frac{q_z(z|z')}{q_z(z'|z)}-\frac{1}{4}\sum_{x,x'}J_x(x|x')\log\frac{q_x(x|x')}{q_x(x'|x)}-\frac{1}{4}\sum_{s,s'}J_s(s|s')\log\frac{q_s(s|s')}{q_s(s'|s)}\nonumber\\
 &=&\frac{1}{2}(R_z-R_x-R_s).
\end{eqnarray}

We note that $I_B(X,S)$ intimated related to the EPRs. This builds up a bridge between these EPRs and irreversible part of the mutual information. Moreover, we also have
\begin{eqnarray}
 \begin{cases}
   R_z=R_x+R_s+ 2I_B(X,S)\ge 0,\\
   R_x+R_s\ge -2 I_B(X,S),\\
   R_z \ge 2I_B(X,S).
 \end{cases}
\end{eqnarray}
This indicates that the time-irreversible MIR contributes to the detailed EPR. In other words, The differences of entropy production rate of the whole system and subsystems provides the origin of the time irreversible part of the mutual information. This gives the nonequilibrium thermodynamic origin of the irreversible mutual information or correlations. Of course, since the EPR is related to the flux directly as seem from above definitions, the origin of the EPR or nonequilibrium thermodynamics is from the non-vanishing information flux for the nonequilibrium dynamics. On the other hand, irreversible part of the mutual information measures the correlations and contributes to the correlated part of the EPR between the subsystems.

\section{A Simple Case: Blind Demon}
A two-state system is connected to two information baths $a$ and $b$. The states of the system are denoted by $\mathcal{X}=\{0,1\}$ respectively. Each bath sends an instruction to the system. If the system adopts one of them, it then follows the instruction and makes change of the state. The instructions generated from one bath are independent, and identically distributed. The probability distributions of the instructions corresponding to the baths read $\{\epsilon_a(0), \epsilon_a(1)\}$ and $\{\epsilon_b(0), \epsilon_b(1)\}$ respectively. Since the system cannot execute two instructions simultaneously, there exists an information demon that makes choices for the system. The demon is blind to care about the system and it makes choices independent, and identically distributed. The choices of the demon are denoted by $\mathcal{S}=\{a,b\}$ respectively. The probability distribution of demon's choices reads $\{P(a)=p, P(b)=1-p\}$. Still, we use $Z=(X,S)$ with $X\in\mathcal{X}$ and $S\in\mathcal{S}$ to denote the joint chain of the system and the demon.

The transition probabilities of the system read
\begin{eqnarray*}
 q_x(x|x')=p\epsilon_a(x)+(1-p)\epsilon_b(x).
\end{eqnarray*}
The transition probabilities of the demon read
\begin{eqnarray*}
 q_s(s|s')=P(s).
\end{eqnarray*}
And the transition probabilities of the joint chain read
\begin{eqnarray*}
 q_z(x,s|x',s')=P(s)\epsilon_{s'}(x).
\end{eqnarray*}
We have the corresponding steady state distributions or the information landscape as,
\begin{eqnarray*}
 \begin{cases}
   \pi_x(x)=p\epsilon_a(x)+(1-p)\epsilon_b(x),\\
   \pi_s(s)=P(s),\\
   \pi_z(x,s)=P(s)\pi_x(x).
 \end{cases}
\end{eqnarray*}
We obtain the information fluxes as,
\begin{eqnarray*}
 \begin{cases}
   J_x(x'\to x)=0, \text{ for all } x,x'\in\mathcal{X} \\
   J_s(s'\to s)=0, \text{ for all } s,s'\in\mathcal{S}\\
   J_z((x',s')\to (x,s))=P(s)P(s')(\pi_x(x')\epsilon_{s'}(x)-\pi_x(x)\epsilon_{s}(x')).
 \end{cases}
\end{eqnarray*}
We obtain the EPRs as
\begin{eqnarray*}
 \begin{cases}
   R_x=0, \\
   R_s=0,\\
   R_z=\sum_xp(1-p)(\epsilon_a(x)-\epsilon_b(x))(\log\epsilon_a(x)-\log\epsilon_b(x) ).
 \end{cases}
\end{eqnarray*}
We evaluate the MIR as
\begin{eqnarray*}
 I(X,S)=-\sum_x\pi_x(x)\log\pi_x(x)+p\sum_x\epsilon_a(x)\log\epsilon_a(x)+(1-p)\sum_x\epsilon_b(x)\log\epsilon_b(x).
\end{eqnarray*}
The time-irreversible part of $I(X,S)$ reads,
\begin{eqnarray*}
 I_B(X,S)=\frac{1}{2}R_z.
\end{eqnarray*}

\section{Conclusion}

In this work, we identify the driving forces for the information system dynamics. We show that the information system dynamics is determined by both the information landscape and information flux representing the time reversible and time irreversible part of the information dynamics. We further demonstrate that the mutual information representing the correlations can be decomposed into time reversible part and time irreversible part originated from the landscape and flux decomposition of the information dynamics. Finally we uncover the intimate relationship between the difference of the entropy production of the whole system and the subsystems and the time irreversible part of the mutual information. This will help for understanding the non-equilibrium behaviour of the interacting information system dynamics in random environments. Furthermore, we believe that our conclusion can be made more general for the BMC with non-Markovian marginal chains which we will discuss in a separate work.

\section*{Acknowledgement} This work was support in part by National Natural Science Foundation of
China (NSFC-21190040, NSFC-11174105, NSFC-91227114, NSFC-91430217) and NSF-PHY-76066 (USA).

\section{Appendix}

Here, we derive the exact form of Mutual Information Rate (MIR, Eq.(13)) in steady state by using the cumulant-generating function.

We write arbitrary time sequence of $Z$ in time $T$ in the form as following
\begin{align*}
  Z^T=\{Z(1),...,Z(i),...,Z(T)\}, \text{ for }T\ge 2,
\end{align*}
where $Z(i)$ (for $i\ge1$) denotes the state at time $i$. The corresponding probability of $Z^T$ is in the following form
\begin{align*}
  P(Z^T)=\pi_z(Z_1)\left\{\prod_{i=1}^{T-1}q_z(Z_{i+1}|Z_i)\right\}.\tag{A.1}
\end{align*}

We let the chain $U=(X,S)$ to denote a process that $X$ and $S$ follow the same Markov dynamics in $Z$ but are independent of each other. Then we have the transition probabilities of $U$ read
\begin{align*}
  q_u(u|u')=q(x,s|x',s')=q_x(x|x')q_s(s|s').\tag{A.2}
\end{align*}
Then the probability of a time sequence of $U$, $U^T$, with the same trajectory of $Z^T$ reads
 \begin{align*}
  P(U^T)=\pi_u(Z_1)\left\{\prod_{i=1}^{T-1}q_u(Z_{i+1}|Z_i)\right\},\tag{A.3}
\end{align*}
with $\pi_u(x,s)=\pi_x(x)\pi_s(s)$ being the stationary probability of $U$.

For evaluating the exact form of MIR, we introduce the cumulant-generating function of the random variable $\log\frac{P(Z^T)}{P(U^T)}$,
\begin{align*}
  K(m,T)=\log\bigg\langle \exp{\left(m\log\frac{P(Z^T)}{P(U^T)}\right)}\bigg\rangle_{Z^T}.\tag{A.4}
\end{align*}
We can see that
\begin{align*}
  &\lim_{T\to \infty}\lim_{m\to 0}\frac{1}{T}\frac{\partial K(m,T)}{\partial m}\\
  &=\lim_{T\to \infty}\frac{1}{T}\bigg\langle \log\frac{P(Z^T)}{P(U^T)}\bigg\rangle_{Z^T}\\
  &=I(X,S).\tag{A.5}
\end{align*}
Thus, our idea is to evaluate $K(m,T)$ at first. We have
\begin{align*}
  K(m,T)&=\log\bigg\langle \exp{\left(m\log\frac{P(Z^T)}{P(U^T)}\right)}\bigg\rangle_{Z^T}\\
  &=\log\left\{\sum_{Z^T}\frac{(P(Z^T))^{m+1}}{(P(U^T))^m}\right\}\\
  &=\log\left\{\sum_{\{Z(0),Z(1),...,Z(T)\}}\frac{(\pi_z^{m+1}(Z_0))}{(\pi_u^{m}(Z_0))}\prod_{i=0}^{T-1}\frac{q_z^{m+1}(Z_{i+1}|Z_i)}{q_u^m(Z_{i+1}|Z_i)}\right\},\tag{A.6}
\end{align*}
where we realize that the last equality can be rewritten in the form of matrices multiplication.

We introduce the following matrices and vectors for Eq. (A.6) such that
\begin{align*}
&\pmb{Q}_z=\left\{(\pmb{Q_z})_{(z,z')}=q_z(z|z'), \text{ for }z,z'\in\mathcal{Z}\right\},\\
&\pmb{G}(m)=\left\{(\pmb{G}(m))_{(z,z')}=\frac{q_z^{m+1}(z|z')}{q_u^m(z|z')} , \text{ for }z,z'\in\mathcal{Z}\right\},\\
&\pmb{\pi}_z=\left\{(\pmb{\pi}_z)_z=\pi_z(z), \text{ for }z\in\mathcal{Z}\right\},\\
&\pmb{v}(m)=\left\{(\pmb{v}(m))_z=\frac{\pi_z^{m+1}(z)}{\pi_u^{m}(z)}\right\},\tag{A.7}
\end{align*}
where $\pmb{Q}_z$ is the transition matrix of $Z$; $\pmb{\pi}_z$ is the stationary distribution of $Z$. It can be also verified that
\begin{align*}
&\pmb{Q}_z=\pmb{G}(0),\\
&\pmb{\pi}_z=\pmb{v}(0),\\
&\pmb{\pi}_z=\pmb{Q}_z\pmb{\pi}_z,\\
&\pmb{1}^{\dagger}\pmb{Q}_z=\pmb{1}^{\dagger},\\
&\lim_{m\to 0}\frac{d\pmb{G}(m)}{dm}=\left\{\left(\lim_{m\to 0}\frac{d\pmb{G}(m)}{dm}\right)_{(z,z')}=q_z(z|z')\log\frac{q_z(z|z')}{q_u(z|z')} , \text{ for }z,z'\in\mathcal{Z}\right\},\\
&\lim_{m\to 0}\frac{d\pmb{v}(m)}{dm}=\left\{\left(\lim_{m\to 0}\frac{d\pmb{v}(m)}{dm}\right)_{z}=\pi_z(z)\log\frac{\pi_z(z)}{\pi_u(z)} , \text{ for }z\in\mathcal{Z}\right\},\tag{A.8}
\end{align*}
where $\pmb{1}^{\dagger}$ is the vector of all 1's with appropriate dimension.

Then $K(m,T)$ can be rewritten in a compact form such that
\begin{align*}
  K(m,T)=\log\left\{\pmb{1}^{\dagger}\pmb{G}^{T-1}(m)\pmb{v}(m)\right\}.\tag{A.9}
\end{align*}

Then, we substitute Eq. (A.9) into Eq. (A.5) and have
\begin{align*}
  I(X,S)&=\lim_{T\to \infty}\lim_{m\to 0}\frac{1}{T}\frac{\partial K(m,T)}{\partial m}\\
  &=\lim_{T\to \infty}\lim_{m\to 0}\frac{1}{T}\frac{\partial \log\left\{\pmb{1}^{\dagger}\pmb{G}^{T-1}(m)\pmb{v}(m)\right\}}{\partial m}\\
  &=\lim_{T\to \infty}\lim_{m\to 0}\frac{1}{T}\left\{(T-1)\pmb{1}^{\dagger}\pmb{G}^{T-2}(m)\frac{d\pmb{G}(m)}{dm}\pmb{v}(m)+\pmb{1}^{\dagger}\pmb{G}^{T-1}(m)\frac{d\pmb{v}(m)}{dm}\right\}\\
  &=\lim_{T\to \infty}\frac{1}{T}\left\{(T-1)\pmb{1}^{\dagger}\pmb{G}^{T-2}(0)\left(\lim_{m\to 0}\frac{d\pmb{G}(m)}{dm}\right)\pmb{v}(0)+\pmb{1}^{\dagger}\pmb{G}^{T-1}(0)\left(\lim_{m\to 0}\frac{d\pmb{v}(m)}{dm}\right)\right\}.\tag{A.10}
\end{align*}
By noting Eq. (A.8) and $T\ge 2$, we obtain Eq. (13) from Eq. (A.10) that
\begin{align*}
  I(X,S)&=\lim_{T\to \infty}\frac{1}{T}\left\{(T-1)\pmb{1}^{\dagger}\pmb{G}^{T-2}(0)\left(\lim_{m\to 0}\frac{d\pmb{G}(m)}{dm}\right)\pmb{v}(0)+\pmb{1}^{\dagger}\pmb{G}^{T-1}(0)\left(\lim_{m\to 0}\frac{d\pmb{v}(m)}{dm}\right)\right\}\\
  &=\lim_{T\to \infty}\left\{\left(1-\frac{1}{T}\right)\pmb{1}^{\dagger}\left(\lim_{m\to 0}\frac{d\pmb{G}(m)}{dm}\right)\pmb{\pi}_z+\frac{1}{T}\pmb{1}^{\dagger}\left(\lim_{m\to 0}\frac{d\pmb{v}(m)}{dm}\right)\right\}\\
  &=\pmb{1}^{\dagger}\left(\lim_{m\to 0}\frac{d\pmb{G}(m)}{dm}\right)\pmb{\pi}_z\\
  &=\sum_{(x,s),(x',s')}\pi_z(x',s')q_z(x,s|x',s')\log\frac{q_z(x,s|x',s')}{q_x(x|x')q_s(s|s')}.\tag{A.11}
\end{align*}

\end{document}